\documentclass[journal,final]{IEEEtran}
\usepackage[utf8]{inputenc}
\usepackage[T1]{fontenc}
\usepackage{graphicx}
\usepackage{grffile}
\usepackage{longtable}
\usepackage{wrapfig}
\usepackage{rotating}
\usepackage[normalem]{ulem}
\usepackage{amsmath}
\usepackage{textcomp}
\usepackage{amssymb}
\usepackage{capt-of}
\usepackage{hyperref}
\usepackage[frozencache=true]{minted}
\usepackage{bytefield}
\usepackage{fancyhdr}
\usepackage{pdfpages}
\usepackage{tabulary}
\usepackage{multirow}
\usepackage[inkscapelatex=false]{svg}
\usepackage{amsmath}
\usepackage{algpseudocode}
\usepackage{url}
\usepackage[disable]{todonotes}
\usepackage[numbers]{natbib}
\usepackage{siunitx}
\usepackage[caption=false,font=footnotesize]{subfig}
\usepackage{booktabs}
\usepackage{xcolor}
\setuptodonotes{inline, color=green!30, shadow}
\hypersetup{
  colorlinks=true,
  citecolor=black,
  linkcolor=black
}
\renewcommand{\arraystretch}{1.2}
\newcommand\copyrighttext{%
  \footnotesize This work has been submitted to the IEEE for possible publication. Copyright may be transferred without notice, after which this version may no longer be accessible.}
\newcommand\copyrightnotice{%
  \begin{tikzpicture}[remember picture,overlay]
    \node[anchor=south,yshift=10pt] at (current page.south) {\fbox{\parbox{\dimexpr\textwidth-\fboxsep-\fboxrule\relax}{\copyrighttext}}};
  \end{tikzpicture}%
}

\author{Manuel~Eggimann, \IEEEmembership{Graduate Student Member,~IEEE,} Abbas~Rahimi, Luca~Benini, \IEEEmembership{Fellow,~IEEE}}
\date{\today}
\title{A 5\,\si{\micro\watt} Standard Cell Memory-based Configurable Hyperdimensional Computing Accelerator for Always-on Smart Sensing}
\hypersetup{
  pdftitle={A Highly Configurable SCM Based HD-Computing Accelerator for Always-On Smart Sensing},
  pdfkeywords={},
  pdfsubject={},
  pdfcreator={Emacs 26.3 (Org mode 9.3.6)}, 
  pdflang={English}}
\begin{document}
\bstctlcite{customize:BSTcontrol}
\maketitle
\begin{abstract}
Hyperdimensional computing (HDC) is a brain-inspired computing paradigm based on high-dimensional holistic representations of vectors. It recently gained attention for embedded smart sensing due to its inherent error-resiliency and suitability to highly parallel hardware implementations.
In this work, we propose a programmable all-digital CMOS implementation of a fully autonomous HDC accelerator for always-on classification in energy-constrained sensor nodes.
By using energy-efficient standard cell memory (SCM), the design is easily cross-technology mappable. It achieves
extremely low power,  5\,\si{\micro\watt} in typical applications, and an energy-efficiency improvement over the state-of-the-art (SoA) digital architectures of up to 3$\times$ in post-layout simulations for always-on wearable tasks such as EMG gesture recognition.
As part of the accelerator's architecture, we introduce novel hardware-friendly embodiments of common HDC-algorithmic primitives, which results in 3.3$\times$ technology scaled area reduction over the SoA, achieving the same accuracy levels in all examined targets.
The proposed architecture also has a fully configurable datapath using microcode optimized for HDC stored on an integrated SCM based configuration memory, making the design ``general-purpose'' in terms of HDC algorithm flexibility.
This flexibility allows usage of the accelerator across novel HDC tasks, for instance, a newly designed HDC applied to the task of ball bearing fault detection.
\end{abstract}

\copyrightnotice
\section{Introduction}
\label{sec:introduction}
\IEEEPARstart{E}{nergy} boundedness is the key design metric and constraint in the development of internet-of-things (IoT) devices~\cite{Chatterjee2019,Bagchi2020,Newell2019}. With more and more sensor modalities integrated into IoT end nodes, the amount of data to process, and the complexity of the processing pipeline increases.
Aiming for uninterrupted operation for years or even indefinitely within the tight power envelope of small batteries or environmental energy harvesting urges to drastically reduce the average power consumption of the sensor nodes themselves.
\IEEEpubidadjcol


Observing that the majority of power consumption in today's wireless sensor devices is spent in data transmission~\cite{Shnayder2004} promotes moving data processing closer to the sensor.
Instead of raw data transmission and centralized processing in the cloud, the data is processed continuously on these so-called \emph{smart sensor} devices~\cite{Spencer2004}. Only the analyzed portion of the information is transmitted (e.g., transmission of a single \texttt{imminent machine failure} message instead of the raw vibration and temperature data).
This cannot be achieved by application-specific integrated circuit (ASIC) designs for deep neural networks alone because \emph{general purpose} always-on smart sensing systems operate in the \si{\micro\watt} range.
Therefore, the next evolution step towards fully self-sustainable always-on smart sensors requires the exploration of new avenues of hardware-software co-design and outside the realm of traditional von Neumann based computing~\cite{Verma2019a, Sebastian2020, Karunaratne2020}.

An energy proportional sensor data processing scheme, where a wake-up circuit (WuC) detects patterns of interest and aggressively duty cycles other circuitry is a viable solution to drastically reduce average power consumption \cite{Miro-Panades2020, Ma2020}.
While there are numerous WuCs, e.g., for biosignal anomaly detection, sound/keyword spotting, incoming radio transmissions in the \si{\micro\watt} range, all of these solutions are highly application-specific.
Considering the cost of custom silicon development and the rapidly widening range of application targets, there is a need for configurable and application-agnostic WuCs with more flexible pattern extraction capabilities than the simple threshold-based solutions, which can suffer from high false-positive rate and thus energy losses of unnecessary wakeups.
\IEEEpubidadjcol

Hyperdimensional computing (HDC) is a brain-inspired computing paradigm that excels in the learning curve, computational complexity of the training, and simplicity of operations for hardware. This makes it a perfect fit for energy-constrained inference applications, and, more specifically for general-purpose always-on sensing~\cite{Ge22, Rahimi2016, Montagna2018}.

In this work, we present the following contributions:
\begin{itemize}
\item We propose a novel flexible and highly energy-efficient all-digital HDC architecture for always-on smart sensing applications achieving up to 3$\times$ higher energy efficiency (191\,\si{\nano\joule/inference}) over the SoA.
\item As part of the architecture, we introduce novel hardware-friendly embodiments of common HDC operators resulting in 3.3$\times$ technology scaled area reduction. \item We provide an evaluation of latch-based associative memories at sub nominal supply voltage conditions in post-layout simulation indicating the potential of at least 3.5$\times$ energy efficiency improvement compared to an SRAM based digital solution.
\item We provide practical application case studies of our approach, including the first investigation (to the best of our knowledge) on the feasibility of HDC for the task of ball bearing fault detection.
\item Finally, using an all-digital approach enables us to publicly release our architecture under the permissive solderpad open-source license\footnote{Will be made available under \href{https://github.com/pulp-platform/hypnos}{https://github.com/pulp-platform/hypnos}}.
\end{itemize}

The remainder of this paper is structured as follows. In Section~\ref{sec:related-work} we elaborate on previous work in the domain of HDC accelerators and always-on classification circuitry and highlight the distinctive novel characteristics of the proposed approach.
Section~\ref{sec:progr-hdc-accel} analyzes in detail the modules of the proposed architecture.
We continue with post-layout analysis on power and area of the design in different target technologies and different design parameter combinations in Section~\ref{sec:impl-results} before we conduct an energy-efficiency and accuracy analysis for several always-on cognitive sensing scenarios in Section~\ref{sec:appl-comp}.
Finally, we conclude in Section~\ref{sec:conclusion}.







\section{Related Work}
\label{sec:related-work}
\begin{table*}
  \centering
  \resizebox*{\linewidth}{!}{
  \begin{tabular}{lccccc|ccc}
    \toprule
    \multicolumn{5}{c}{\emph{Application Specific}} & \multicolumn{4}{c}{\emph{General Purpose}} \\
                                            & \citeauthor{Cho2019} \cite{Cho2019}   & \citeauthor{Giraldo2020} \cite{Giraldo2020} & \citeauthor{Shan2020} \cite{Shan2020} & \citeauthor{Zhao2020} \cite{Zhao2020} & \citeauthor{Wang2020a} \cite{Wang2020a} & \citeauthor{Miro-Panades2020} \cite{Miro-Panades2020} & \citeauthor{Rovere2018} \cite{Rovere2018} & \bf{This Work}                                                  \\
    \midrule                                                                                                                                                                                                                                                                                    
    Applications                            & VAD                                   & Keyword Spott.                              & Keyword Spott.                        & EMG                                   & Slope Matching                          & Wake-up Radio, Interrupts                             & General Purpose                           & General Purpose                                                 \\ 
    Technology                              & 180nm                                 & 65nm                                        & 28nm                                  & 180nm                                 & 180nm                                   & 28nm                                                  & 130nm                                     & 65nm / 22nm                                                     \\
    Cross Tech.                             & Low                                   & Medium                                      & High                                  & Low                                   & Low                                     & Low                                                   & Low                                       & High                                                            \\
    Power Envelope                          & \textasciitilde{}\SI{10}{\micro\watt} & \textasciitilde{}\SI{10}{\micro\watt}       & \textasciitilde\SI{500}{\nano\watt}   & \textasciitilde\SI{13}{\micro\watt}   & \textasciitilde\SI{75}{\nano\watt}      & \textasciitilde\SI{33}{\micro\watt}                   & \textasciitilde\SI{2.2}{\micro\watt}      & max. \textasciitilde\SI{25}{\micro\watt}, typ. \textasciitilde\SI{5}{\micro\watt}\\
    Classification Scheme                   & NN                                    & MFCC, LSTM                                  & DSCNN                                 & ANN                                   & Threshold, Slope                        & -                                                     & Threshold Sequence                        & HDC                                                             \\
    Configurability                         & App. Specific                         & App. Specific                               & App. Specific                         & App. Specific                         & Limited                                 & App. Specific                                         & Medium                                    & High                                                            \\
    Area                                    & \SI{15.6}{\square\milli\meter}        & \SI{2.56}{\square\milli\meter}              & \SI{0.23}{\square\milli\meter}        & \SI{0.925}{\square\milli\meter}       & \SI{2.5}{\square\milli\meter}           & \SI{4.5}{\square\milli\meter}                         & \SI{0.054}{\square\milli\meter}           & \SI{1.43}{\square\milli\meter} / \SI{0.29}{\square\milli\meter} \\
  \end{tabular}
  }
  \caption{Comparison of state-of-the-art WuCs with our proposed HDC based WuC.
    Area numbers are reported 65nm and 22nm technology while power
    consumption is reported in 22nm for a compute intensive language classification algorithm and
    a typical always-on classification algorithm for EMG data. }
  \label{tab:wuc-comparison}
\end{table*}
Tackling the power-consumption challenge of always-on sensing in a hierarchical manner using WuCs to apply aggressive duty cycling on more involved data processing modules is not a new idea.
In the recent past, there have been several publications on low-power always-on wake-up circuitry in various domains.
Table \ref{tab:wuc-comparison} gives an exemplary overview of current wake-up circuitry research using selected publications in the recent past.
Keyword spotting and voice activity detection (VAD) is a very actively researched target for always-on sensing; \citeauthor{Giraldo2020} present a low power WuC for speech detection, speaker identification, and keyword spotting with integrated preprocessing blocks for MFCC generation and LSTM accelerator for classification \cite{Giraldo2020}.
\citeauthor{Shan2020} proposed another implementation in the same application domain with state-of-the-art energy efficiency on the task of two-word keyword spotting using binarized depth-wise separable CNN's operating at near-threshold \cite{Shan2020}.
At the lower end of the power consumption spectrum \citeauthor{Cho2019} present a \SI{142}{\nano\watt} VAD circuitry with integrated analog-frontend that combines a configurable always-on time-interleaved mixer architecture with a heavily duty cycled neural-network processor \cite{Cho2019}.

Monitoring life signals is another very active field;
In the context of cardiac arrhythmia detection, \citeauthor{Zhao2020} combine a level-crossing ADC with asynchronous QRS-complex detection circuitry with an artificial neural network accelerator to benefit from the energy advantage of non-Nyquist sampling \cite{Zhao2020}.
Although these solutions achieve outstanding energy efficiency in their particular application domain, they are hardwired for the respective task.

More in line with the target of a flexible and configurable smart sensing platform are \citeauthor{Miro-Panades2020}; They present an asynchronous RISC processor with an integrated wake-up radio receiver for efficient low-latency wake-up from several external and internal triggers.
While their architecture achieves outstanding reaction time to interrupts without the need for a high-frequency clock, the wake-up circuitry lacks the interface and compute capability to perform actual data processing for data input pattern dependent wake-up \cite{Miro-Panades2020}.
\citeauthor{Wang2020a} present a configurable WuC resembling the work in \cite{Zhao2020} that combines an LC-ADC with a set of asynchronous detector blocks to extract low-level signal properties like peak amplitude, slope, or time interval between peaks.
Each detector can be configured with a threshold, and the individual detector output can be logically fused to a single wake up signal.
Although their architecture uses minimal power, continuous detection of more complex patterns is entirely outside the capabilities of a detector-set approach \cite{Wang2020a}.

To the authors' knowledge, the only low power WuC with slightly more sophisticated pattern matching capabilities was introduced by \citeauthor{Rovere2018}.
Instead of analyzing the delta-encoded signal from the LC-ADC with hardwired detectors, they continuously match the input signal against a sequence of upper and lower amplitude thresholds with up to 16 threshold segments.
This scheme equates to matching the input signal's approximate amplitude slope against a configurable pre-trained prototypical signal slope of interest \cite{Rovere2018}.
Their approach proved successful for pathological ECG classification and binary hand gesture recognition (finger-snap or hand clapping).
Still, detecting more complex patterns in the spatial or time dimension remains outside their proposed architecture's scope.

Hyperdimensional computing (HDC) is an energy-efficient and flexible computing paradigm for near-sensor classification that gracefully degrades in the presence of bit errors, and noise~\cite{Kanerva2009, Rahimi2019, Rahimi2017}. Various works showcased HDC's few-shot learning properties and energy efficiency in multiple domains like biosignal processing~\cite{Moin2018, Burrello2019, Chang2019}, language recognition~\cite{Joshi2017}, DNA sequencing~\cite{Imani2018}, or vehicle type classification~\cite{Kleyko2014}.

In emerging hardware implementations, the HDC's inherent error-resiliency is leveraged for novel non-volatile memory (NVM) based in-memory computing architectures~\cite{Karunaratne2020, Wu2018a, Li2016}. Targeting FPGAs, efficient mappings of binary and bipolar HDC operations are proposed~\cite{Schmuck2019,Salamat2019, Salamat2020}. However, the only complete digital CMOS-based HDC accelerator was recently introduced by \citeauthor{Datta2019}. They propose a data processing unit (DPU) based encoder design that interconnects with a ROM based item memory, and a fully parallel associative memory~\cite{Datta2019}.
While their implementation indeed excels in throughput, its' configurability as well as area- and energy-efficiency are limited;
Their encoder architecture is restricted to what they call \emph{generic} multi-stage HDC algorithms with a hardwired encoder depth in feedforward configuration imposing hard limits on the supported encoding schemes.
From an energy-efficiency and area standpoint, their design suffers a lot from using a large read-only-memory (ROM) for item memory (IM) and pipeline registers in the very wide datapath of every encoding layer.

Our proposed architecture targets the sub 25\si{\micro\watt} power envelope (resulting in a lifetime of about four years from a small lithium-thionyl chloride coin cell battery).
The always-on smart sensing circuitry leverages the flexibility of HDC to perform energy-efficient end-to-end classification on a diverse set of input signals.
We achieve higher configurability, a reduction of 3.1$\times$ in area and up to 3.3$\times$ improvement in energy-efficiency than the current SoA in HDC acceleration and present a first-in-class flexible and technology agnostic digital CMOS architecture for near sensor smart sensing wake-up circuitry.

\section{Programmable HDC-Accelerator Architecture}
\label{sec:progr-hdc-accel}
\subsection{Hyperdimensional Computing}
\label{sec:primer-hdc-theory}
Hyperdimensional Computing (HDC) or vector symbolic architectures (VSAs) in general, is a brain-inspired compute paradigm that recently is gaining attention \cite{Kanerva2009}.
Its core idea is to map low-dimensional input data, i.e., raw sensor data or features thereof, to vectors of very high dimensionality (cardinality in the order of thousands).
The procedure of input to HD space mapping is commonly called \emph{hyperdimensional encoding}.
HDC defines simple operations on vectors to aggregate their informational content into a single vector.
\emph{Binding} a vector $V_a$ to another vector $V_b$ creates a vector that is dissimilar to both inputs and thus may be used to represents the mapping $V_a:V_b$.
\emph{Bundling} several input vectors yields a vector most similar to all of its inputs, therefore representing the set of its input vector.
The unary \emph{Permutation} operation maps a single vector deterministically to an entirely unrelated subspace.
Combining these three operations on multiple channels or a time-sequence of mapped input vectors (using a so-called \emph{item memory}) captures high-level signal characteristics of the underlying data in an error-resilient and flexible manner \cite{Imani2017a}.

The inverse mapping of HD Vectors to the low dimensional output space, i.e., the index of a classification result, is enabled by the \emph{Associative Lookup} operation.
This operation finds the most similar vector to the input within a set of stored HD vectors.
There are various embodiment options for VSAs, differing in the concrete representation of the individual dimensions and actual implementations of \emph{Binding}, \emph{Bundling} and the similarity metric.
In this work, we concentrate on the so-called binary spatter code (BSC), a digital CMOS friendly VSA that uses a single bit per dimension.
BSC uses XOR for the binding and majority vote for the bundling operation with Hamming distance as the implied similarity metric for associative lookups.

\begin{figure}
    \centering
    \includesvg[width=\linewidth]{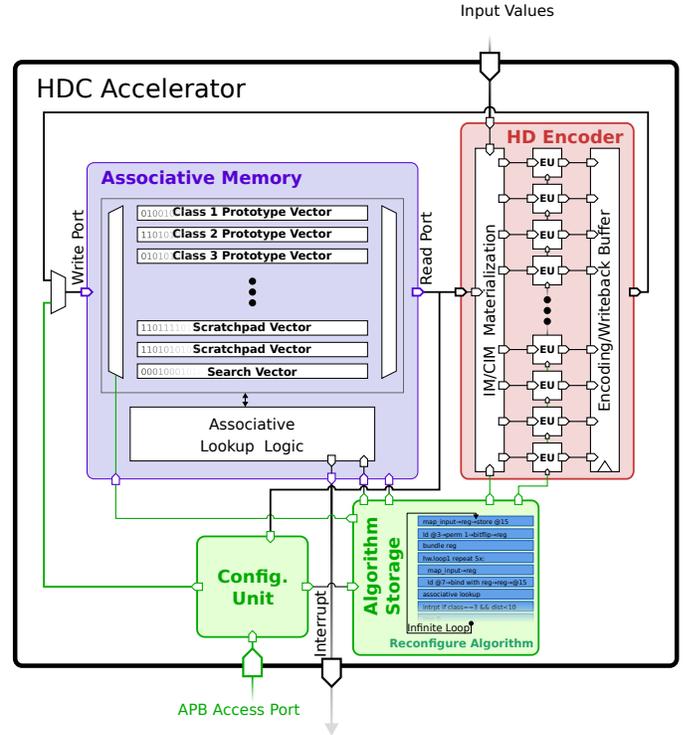}
    \caption{High-level structure of the proposed HDC accelerator. The associative memory (blue) is responsible for storage and associative lookup of prototype vectors and serves as a scratchpad memory for the HD Encoder (red). Encoder and associative memory are orchestrated by user programmable algorithm storage (green).}
    \label{fig:arch_overview}
\end{figure}
\begin{figure*}  
  \centering
  \subfloat[]{
    \includesvg[width=0.6\linewidth]{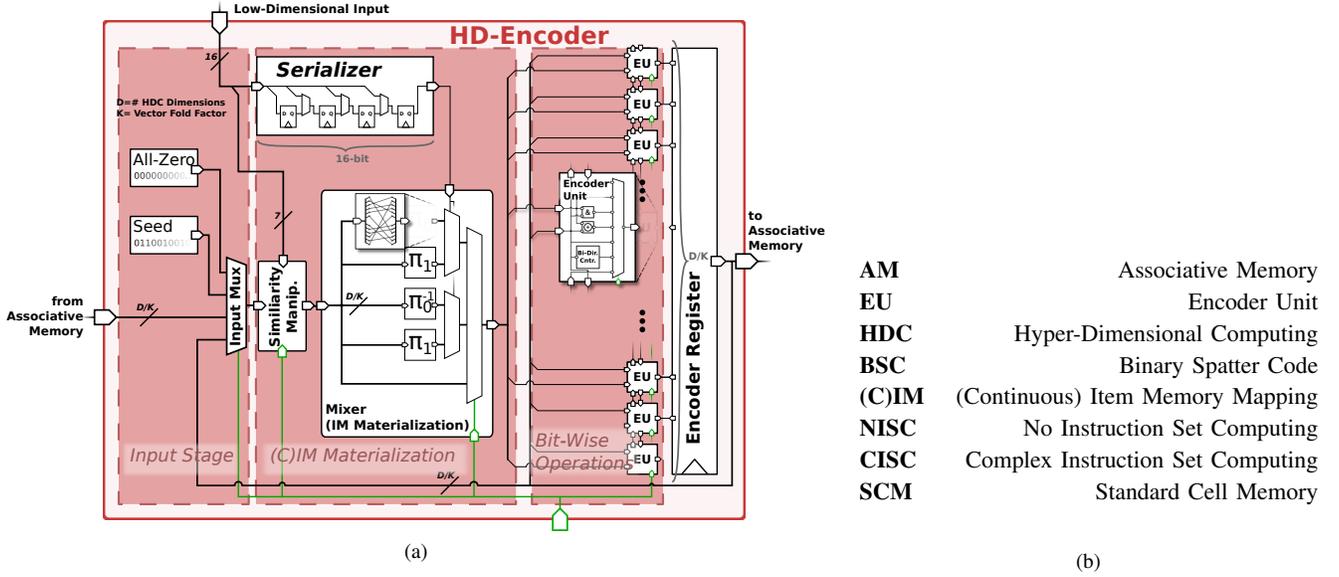}
    \label{fig:hd_encoder}
  }
  \subfloat[]{
    \small
    \begin{tabular}[b]{lr}
      \textbf{AM} & Associative Memory \\
      \textbf{EU} & Encoder Unit \\
      \textbf{HDC} & Hyper-Dimensional Computing \\
      \textbf{BSC} & Binary Spatter Code \\
      \textbf{(C)IM} & (Continuous) Item Memory Mapping\\
      \textbf{NISC} & No Instruction Set Computing \\
      \textbf{CISC} & Complex Instruction Set Computing \\
      \textbf{SCM} & Standard Cell Memory \\
      \label{tab:acronyms}
    \end{tabular}
  }
    
  \caption{\protect\subref{fig:hd_encoder} Architecture of the HDC Encoder responsible for (Continouus) Item Memory  materialization and search vector encoding. The width of the datapath is a function of the HDC dimensionality (D) and the design parameter K (discussed in Section~\ref{sec:vector-folding}).
  \protect\subref{tab:acronyms} Table of accelerator related acronyms.}
\end{figure*}

\subsection{Overview}
\label{sec:overview}
Figure \ref{fig:arch_overview} illustrates the three major components of the accelerator, which we describe in detail in the following subsections; the \textbf{associative memory} (AM) stores the prototype vectors and performs the associative lookup operations, the final step of most HDC algorithms.
The \emph{hyperdimensional encoder} (HD-Encoder) is responsible for mapping low-dimensional input values to HD-vectors.
It operates on HD vectors from the AM or its own output in an iterative manner.
The AM and HD-encoder are managed by a small controller circuit that sequentially consumes a stream of compact microcode instructions and accordingly reconfigures the datapath.
A tiny user-programmable configuration memory supplies this microcode stream.
\subsection{HD-Encoder}
\label{sec:hd-encoder}
The first step of every HDC classification algorithm is mapping a dense input space to a high-dimensional holistic representation.
Most current algorithms encode the low-dimensional input data into a single high-dimensional search vector representing the whole or a subset of the input.
The search vector is then compared with prototype vectors stored in the AM that represent the different classes.
The differences between the various HDC algorithms mainly lay in the particular encoding algorithms.
They are crafted to capture relevant characteristics from the raw data, e.g., amplitude distribution, spatial or temporal features, and are highly target application dependent.
Thus, it is mainly the encoder's versatility that affects the affinity of an HDC accelerator for different algorithms. 

Figure \ref{fig:hd_encoder} illustrates our proposed encoder architecture.
It consists of three main components connected in a combinational pipeline.
The input stage of the encoder multiplexes between 4 different input sources; the all-zeros vectors, a hardwired random seed vector, a vector addressed from AM, or the HD-encoder's own output.
The IM materialization stage maps input data to item vectors using either quasi-orthogonal vectors (IM) or continuous item mapping (CIM).
The encoder's final stage are the bitwise encoder units that perform binary or unary operations on the individual bits of the vectors.

There are no pipeline registers in the very wide datapath between the encoder stages.
Although this design choice reduces throughput, it increases the energy efficiency of our architecture.

\subsubsection{Encoder Units}
\label{sec:encoder-units}
The Encoder Unit processes one dimension of the input vector.
Besides the combinational logic for the binary and unary bitwise operations, each unit contains an output Flip-Flop that stores the result after each encoding cycle.

Additionally, there is one saturating bidirectional 5-bit counter per unit to perform the bundling operations.
Analyses in \cite{Schmuck2019} showed that for dimensions up to 10000, a 5-bit saturating counter implementation still achieves the same bundling capacity as a full precision model.

A noteworthy detail of the saturating counter is its possibility to evict the current counter value to the AM in a bit-serial manner (i.e., one cycle for each of the five counter bits).
Eviction and loading of the counter state allow the proposed design to execute multistage encoding algorithms with nested bundling operations.

\subsubsection{Mixing Stage}
\label{sec:mixing-stage}
The Mixer submodule visualized in figure \ref{fig:hd_encoder} generates quasi-orthogonal pseudo-random HD vectors.
The rematerialization, i.e., on-the-fly regeneration, of such vectors is an area-efficient alternative to explicit storage of large numbers of item vectors required for input to HD space mapping.

The mixer stage feeds the input vector selected by the encoder input stage through one of two hardwired random permutations \(\pi_{0}\) and \(\pi_{1}\).
This enables the encoder to map a given low-dimensional binary input datum \(w\) from the input domain \(\mathbb{D}\) to the pseudo-random HD-vector \(V_{w}\) by iteratively applying one of the two permutations to a hardwired random seed HD-vector \(S\):

\begin{equation}\label{eq:im_mapping}
    V_{w}=\prod_{k=0}^{n}\pi_{i}S
    \text{, for } i = \begin{cases}
        0 &, \text{if $w_k=0$}\\
        1 &, \text{if $w_k=1$}\\
    \end{cases}
\end{equation}

where \(w_{k}\) denotes the k\textsuperscript{th} bit position in the input word w's binary representation and \(n=\log_{2}|\mathbb{D}|\).
The resulting HD-vectors \(V_{w}\) are all quasi-orthogonal, given that \(\pi_{0}\) and \(\pi_{1}\) do not commute.

For algorithms that require random access to the item memory, the above scheme rematerializes the item vector with time complexity \(\mathcal{O}(\log_{2}|\mathbb{D}|)\).
However, many algorithms use IM-mapping to bind a value vector \(V_{value}\) to a channel label \(V_{chn[k]}\).
In these scenarios, the channel label vectors are used with a fixed ordering assuming the raw data is feed to the accelerator using a fixed channel ordering.
We can therefore reduce the time complexity to \(\mathcal{O}(1)\) with the mapping:

\begin{equation}
    V_{chn[k]}=\begin{cases}
        S & \text{if $k=0$}\\
        \pi_0V_{chn[k-1]}& \text{if $k>0$}\\
    \end{cases}
\end{equation}

where we store the channel label from the previous iteration in an unused row of the associative memory.

Our proposed IM-mapping approach is more area-efficient than storing random vectors in a large ROM and scales well to large input domains whose cardinality is unknown in advance.
From a hardware perspective, the mixer stage translates to N 4-input and N 2-input multiplexers, where N denotes the datapath width and some moderate wiring overhead caused by the random permutations.

\subsubsection{Vector Folding}
\label{sec:vector-folding}
If synthesized with default parameters, the proposed HD-Accelerator contains a datapath wide enough to process a whole HD-Vector in a single cycle.
However, as will be analyzed in more detail in Chapter \ref{sec:impl-results}, going for a more parallel architecture does not always yield the most energy-efficient solution for a given target technology.
Thus, apart from various other modifiable design parameters, the design exposes the \emph{Vector Fold} parameter; It allows to tune the design for the optimal amount of parallelism to achieve maximum energy efficiency.
Increasing the value of \emph{the vector fold} splits a single $D$-dimensional vector into $K$ smaller subparts of equal size.
The datapath of the accelerator shrinks accordingly and only processes one subpart at a time.
While the throughput of the accelerator at constant frequency decreases by $K$, the area of the HD-Encoder, dominated by the saturating counters, reduces similarly by a factor of $K$.

An important detail is that by decreasing the datapath width, we also reduce the permutations' operand size within the similarity manipulator and the mixing stage.
If we stick with the same IM-Mapping scheme described above, all subparts of a vector would be identical since they all pass through the same hardwired permutations $\pi_0$ and $\pi_1$ of size $\frac{D}{K}$.
The IM mapping scheme in equation \eqref{eq:im_mapping} is thus modified as follows:

\begin{equation}
    V_{w}^*=\pi_{\texttt{idx}}V_{W}
\end{equation}
with,
\begin{equation}
    \pi_{\texttt{idx}}=\prod_{k=0}^{log_{2}(j)}\pi_{i}
    \text{, for } i = \begin{cases}
        0 &, \text{if $h_k=0$}\\
        1 &, \text{if $h_k=1$}\\
    \end{cases}
\end{equation}
where $h$ is the value of the dedicated \emph{part index counter} which holds the index of the current vector subpart.
This yields a unique set of permutations ${\pi_o^*, \pi_1^*}$ per vector part at the expense of $\mathcal{O}(log((K))$ additional mixing cycles.

\subsubsection{Similarity Manipulator}
\label{sec:simil-manip}
\begin{figure}
    \centering
    \includesvg[width=\linewidth]{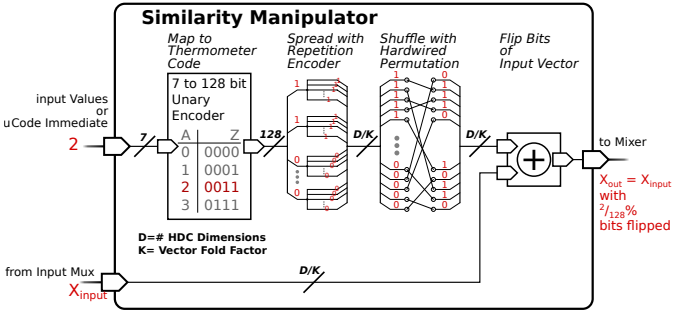}
    \caption{Structure of the Similarity Manipulator stage}
    \label{fig:man_module}
\end{figure}
The Similarity Manipulator stage transforms the mixing stage's output vector by flipping a configurable number of its bits.
This operation is a fundamental building block of various high-level operations like binarized B2B bundling \cite{Schmuck2019}, CIM mapping \cite{Rahimi2019} and exponential forgetting.
Figure \ref{fig:man_module} shows its internal structure; The 7-bit input word \(w\) is first mapped to a 128-bit unary representation \(w_{\text{unary}}\).
This unary representation is spread to the target HD-vector dimensionality \(D/K\) by repeating each bit of \(w_{\text{unary}}\) \(\frac{D}{K\times128}\) times.
The resulting vector passes through a hard-wired random permutation to distribute the 'ones' over all the vector dimensions. The result is XOR-ed with the input vector.
A limitation of the proposed solution is that a uniform distribution of the input words does not yield equal distribution of the probabilities for a bit to be set across the HD-Vector's input dimensions.
A multi-cycle approach can be used for operations where equal bit-flipping probability is a hard requirement; First, a bitmask with the desired bit-density is generated by passing the all-zero vector through the manipulator stage with the input word \(w\).
This mask is subsequently mixed in the Mixing stage using the same input word \(w\) to randomize the position of the 'ones' in the bitmask.
The resulting bitmask is ultimately XOR-ed with the input HD-vector within the encoder units.



\subsection{Fully-Synthesizable Associative Memory}
\label{sec:synthesizable-cam}
For a given search vector, the AM looks up the most similar vector currently stored within the memory.
However, the obvious approach to combine traditional SRAMs to store the HD-vectors with digital logic yields suboptimal results. Although SRAMs are the go-to solution for fast and area-efficient volatile on-chip memory, conventional SRAM macro generators are not optimized for the extremely wide memory aspect ratios needed for parallel access to HD-vectors. Also they are less energy-efficient under low $V_{DD}$ conditions for low bandwidth applications \cite{Teman2016, Meinerzhagen2011}.
The nature of hyperdimensional computing with lots of simple, componentwise operations demands a non-von Neumann scheme of computation with computational logic intermixed with memory cells.

\begin{figure}
  \centering
  \includesvg[width=\linewidth]{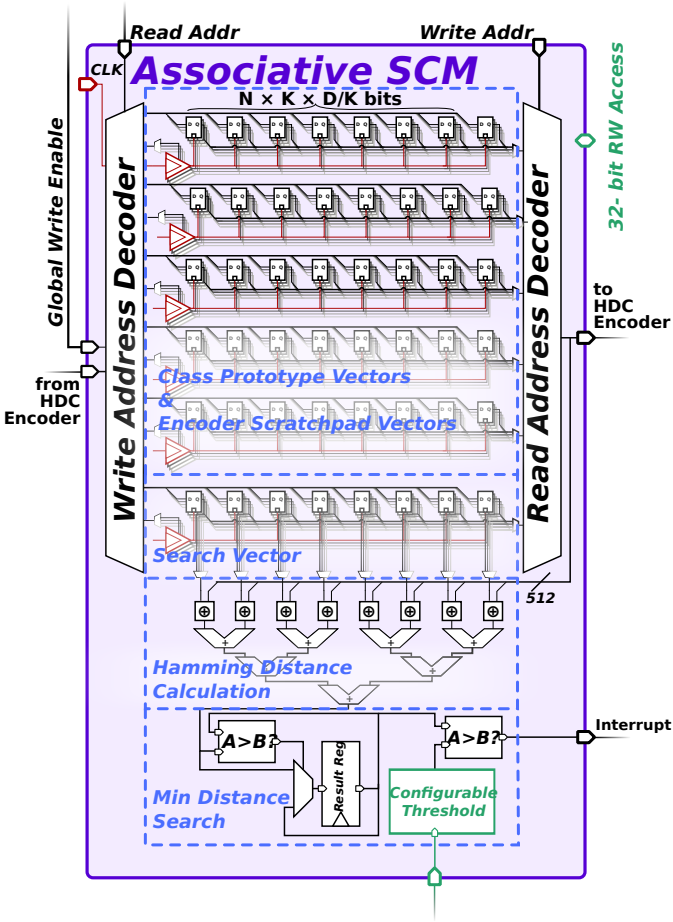}
  \caption{Architecture of the latch-cell based all-digital AM. Vectors can be read and written simultaneously in subrows of length $D/K$. The last vector within the memory acts as the search vector for the associative lookup logic. The $D/K$-bit adder tree for the popcount operation is shared by all memory rows. The distance of the most similar entry is compared with configurable threshold and conditionally raises an interrupt line to an external peripheral (e.g. power management unit in an SoC)}
  \label{fig:cam_architecture}
\end{figure}

\subsubsection{Using Latch Cells as Memory Primitives}
\label{sec:using-latch-cells}
Figure \ref{fig:cam_architecture} shows the structure of the AM in our design; latch cells are used as primitive memory elements instead of flip-flops due to their lower area (~-10\%) and energy (~-20\%) footprint \cite{Meinerzhagen2011}.
Each row of the memory consists of $D/K$ latch cells and a single glitch-free clock gate.
These row clock gates are activated by the one-hot encoded write address.
A two-port design allows fetching a new HD-Vector from AM into the HD-encoder while simultaneously writing back the previous result without any stalls or energy costly pipeline registers in the wide datapath.

In most HDC based classification schemes, the AM is solely keeps hold of the prototype vectors representing the individual classes.
The proposed architecture differs in that regard by using rows of the AM to store the iterative encoding process's intermediate results.
The AM thus serves the double purpose of a register file for entire HD-vectors (or vector subparts in case \emph{vector fold} $K > 1$).

Although latch cells drastically reduce the impact on area footprint compared to flip-flops, their usage can complicate static timing analysis (STA).
Due to their transparent nature during write access, one must take care not to introduce combinational loops.
While \citeauthor{Teman2016} suggest decoupling the memory by using flip-flops at the IO boundary of the memory \cite{Teman2016}, we repurposed the output register in the encoder stage to break combinational loops.
This approach, coupled with multicycle path constraints for STA \cite{Teman2016}, allows treating the AM like a regular flip-flop based synchronous design during synthesis.

\subsubsection{Associative Lookup Logic}
\label{sec:assoc-look-logic}
As can be seen in figure \ref{fig:cam_architecture}, the HD-vector slot acts as the search vector in the proposed architecture.
While we could directly use the write input into the memory as the search word, this would prevent the vector folding feature's usage since our write port would not have a full vector width anymore.
The lookup logic iterates over each memory row, calculating the Hamming distance between one subpart of the search vector and a subpart of one of the stored HD-vectors at a time.
The control logic accumulates the Hamming distance between the subparts and iteratively determines the most similar entry's index and distance.

    
\subsection{An ISA for HD-Computing}
\label{sec:an-isa-hd}
Previously proposed HDC accelerator designs hardwired large portions of their datapath to execute HD-algorithms of a particular structure \cite{Rahimi2019}.
On the other hand, the architecture we are proposing is not bound to execute only one specific class of algorithms.
A control unit continuously reconfigures the datapath according to a stream of microcode instructions fetched from a tiny embedded configuration memory.
This allows the accelerator to be reconfigured at runtime to execute algorithms of a much larger variety by altering the microcode stored in the configuration memory.
After configuration, the algorithm is executed autonomously without any further interaction of a host processor.

We propose a 26-bit instruction set architecture (ISA) with the encoding space split into 25-bit No-instruction-set computing (NISC) and 25-bit Complex-instruction set computing (CISC) instructions.

\subsubsection{Low-level NISC Instructions}
\label{sec:low-level-nisc}
The NISC instructions directly encode the select signals of the multiplexers within the HD-encoder and the address lines of the HD-memory.
Figure \ref{fig:nisc_instruction_bitfield} summarizes the function of the bitfields with a single 25-bit NISC instruction.
They provide fine-grained control over the datapath with the RIDX and WIDX fields acting like source and destination register operands in a conventional ISA.
conventional ISA.
However, since the Encoder unit contains an output Flip-Flop, many vector transformation operations can be performed without AM access using feedback.

If we synthesize the architecture with a \emph{Vector Fold} parameter larger than 1, all instructions only process a smaller subpart of the complete HD-vector.
The control unit does not transparently iterate over all subparts of the vector but leaves control to the user through the \emph{part index counter}.
The counter's value is automatically appended to the read- and write-port address lines of the AM and thus controls which subpart of the HD-vector is affected by the current instruction.
The counter can be cleared, increased, and decreased with dedicated instructions.

The rationale behind leaving control over the subpart iteration scheme to the user is that we also want to support iteration over the vector parts in the outermost loop of an HD-algorithm instead of only iterating in the innermost loop.
That is, instead of first applying a transformation on all subparts of a  vector before switching to the next transformation, we want the possibility to apply all operations of an HD-encoding algorithm on the first subpart and repeat the whole algorithm for subsequent subparts.
For the first iteration scheme, we would have to swap the bundling counters' state after every bundling operation since we do not have individual counters for each vector part.
The second iteration scheme does not require state eviction but requires multiple iterations over the input stream.
\begin{figure}[t]
    \centering
    \begin{bytefield}[bitwidth=2em,endianness=big]{13}
        \bitheader{0,1,2,4,5,6,7,8,9,10,12} \\
        \bitbox{3}{\footnotesize ENCSEL} & \bitbox{1}{\tiny SMEN} & \bitbox{1}{\tiny SMSEL} & \bitbox{1}{\tiny MXEN} & \bitbox{1}{\tiny MXINV} & \bitbox{1}{\tiny MXSEL} & \bitbox{3}{\footnotesize OP} & \bitbox{1}{\tiny BNDEN} & \bitbox{1}{\tiny BNDRST} \\
        \bitheader{0,5,6,11,12} \\
        \bitbox{1}{\tiny WBEN} & \bitbox{6}{\footnotesize RIDX} & \bitbox{6}{\footnotesize WIDX}\\
    \end{bytefield}
    \scriptsize{
    \renewcommand\arraystretch{1.3}
    \begin{tabulary}{\linewidth}{lL}
        \textbf{ENCSEL} & Select between the all-zero vector, a vector from AM and the current HD-encoder output as input for the encoder stage\\
        \textbf{SMEN} & Enable/Bypass Similarity Manipulator Stage \\
        \textbf{SMSEL} & Select between external input data and internal register as input for similarity manipulator stage \\
        \textbf{MXEN} & Enable/Bypass Mixing Stage \\
        \textbf{MXINV} & Select Inverse Permutation set in Mixing Stage \\
        \textbf{MXSEL} & Select between permutation $\pi_0$ and $\pi_1$ or if MXINV is set between $\pi_0^{-1}$ and $\pi_1^{-1}$.\\
        \textbf{OP} & Select operations to be performed in Encoder Units. \\
        \textbf{BNDEN} & Enable bundle counter thus bundling the current encoder output. \\
        \textbf{BNDRST} & Reset the bundle counter to its initial value \\
        \textbf{WBEN} & Enable write back of the encoder output to AM at index WIDX. If disabled HD-encoder is only stored in output buffer. \\
        \textbf{RIDX} & Read index in case vector from AM is used as encoder input. \\
        \textbf{WIDX} & Write index if the result of current iteration is written back to AM (WBEN = 1).\\
    \end{tabulary}
    }
    \caption{NISC Instruction format}
    \label{fig:nisc_instruction_bitfield}
\end{figure}

\subsubsection{CISC Instructions}
The CISC instructions encode multicycle HDC operations and instructions for code size reduction and host interaction.
\paragraph{High-level HDC Operations}
For several HDC transformations, there are dedicated high-level multicycle instructions.
Providing CISC instructions on top of the NISC ISA keeps the number of control signals and thus the instruction with small.
Furthermore, mapping common HDC operations like IM-Mapping or associative lookup to single CISC instructions reduces the given HDC-algorithm's code size.

The \emph{AM\_SEARCH} instruction starts the associative lookup procedure within the AM.
The vector currently stored at the highest index is used as the search vector.
As its only operand, the instruction takes an immediate that limits the search space to a maximum index.
Only vectors stored at an index smaller than the given maximum index are considered during the lookup operation.
The immediate value thus allows partitioning the AM dynamically into scratchpad and prototype memory.

The \emph{MIX} instruction applies multiple mixing cycles to the current content of the encoder register and hence is the basis of IM-mapping.
The mixing value is either an immediate, the current value of the \emph{part index counter} or an externally supplied value, e.g., digital data from a sensor.

\paragraph{Host interaction and Code Size Reduction}
An autonomous WuC requires to conditionally signal a target system about the result of the classification algorithm.
The proposed design uses a dedicated interrupt instruction to conditionally (or unconditionally) assert an interrupt signal line.
The instruction has two operands:
\begin{itemize}
  \item \emph{Similarity Threshold} - The interrupt is not raised if the last associative lookup operation yielded a result with a Hamming distance higher than the given value.
  \item \emph{Index Threshold} - The interrupt signal is not raised if the index of the most similar vector found in the last associative lookup operation is higher than the given threshold.
\end{itemize}
One use case of these thresholds is to wake up the target system only if the HDC classification algorithm detects one particular class with a certainty above a specific threshold.

For the architecture to be autonomous and energy-efficient, the amount of memory required to map a given HD algorithm to the proposed ISA must be kept small.

Thus the algorithm storage in our design supports up to 3 nested hardware loops.
Each loop is initiated with a single instruction containing a 10 bit immediate for the number of iterations and a 10 bit immediate for the instruction address that marks the end of the loop body.

The combination of dedicated instructions for commonly used HDC algorithmic primitives and code size reducing features like hardware loops results in a high expressiveness of the ISA.
All examined HDC algorithms (see Section~\ref{sec:appl-comp}) can be mapped with less than 64 instructions.


\subsubsection{An Example Configuration for Language Recognition}
Language Recognition is a commonly used example application in the field of HDC \cite{Rahimi2016a, Rahimi2017, Joshi2017, Li2016, Wu2018a, Karunaratne2020, Ge22}.
The task is to determine the language given a sentence in the form of a character string.
For a text corpus with 21 European languages, HDC achieves accuracies of up to 96.7\% \cite{Rahimi2016a}.
The algorithm consists of four main steps; In the preprocessing step, the test sentence is split into so-called n-grams, substrings of the test sentence, obtained when applying a sliding window of size $n$ over the character string.
In the next step, the individual n-grams of the sentence are each mapped to an HD-vector according to

$V_{\texttt{n-gram}}=\pi^{n-1}(V_{\texttt{n-1}}) \oplus
\pi^{n-2}(V_{\texttt{n-2}}) \oplus \ldots \oplus V_{\texttt{0}}$

with $V_\texttt{k}$ denoting the HD-vector corresponding to the character at index $k$ within the n-gram.
This vector is obtained through IM mapping using 27 random HD-vectors (26 characters in the Latin alphabet plus one for whitespaces).
$\pi^{k}$ denotes the repeated application of a bit permutation (most commonly a binary shift operation), and $\oplus$ is the bind operator (XOR for BSC).
The n-gram vectors $V_\texttt{n-gram}$ for the test sentence are then bundled together to a single search vector $V_{\texttt{sentence}}$ and in the final step compared with prototype vectors for each language in the AM.
The model of the described algorithm, thus the prototype vectors are obtained by bundling together all sentence vectors $V_{\texttt{sentence}}$ of the training dataset of a language.

In practice, an n-gram size of 4 proved to yield the best performance in terms of accuracy \cite{Rahimi2016a}.

Listing \ref{lst:2} shows the above algorithm for n=4 in Pseudocode;

\begin{listing}[htbp]
\label{lst:2}
  \caption{Pseudo code of an HDC algorithm for language recognition.}
\begin{algorithmic}[1]
    \scriptsize
    \State{$i \leftarrow 0$}
    \State{$\texttt{char\_vec}_{i-[0,1,2,3]} \leftarrow 2048'b0$}
    \State{$\texttt{ngram}_{i-[0,1]}\leftarrow 2048'b0$}

    \For{char in sentence}
    \State{$\texttt{char\_vec}_i \leftarrow im\_map(\texttt{char})$}
    \State{$\texttt{ngram}_i \leftarrow  \pi(\texttt{ngram}_{i-1}) \oplus
      \texttt{char\_vec}_i \oplus \pi^4(\texttt{char\_vec}_{i-4})$}
    \State{$i \leftarrow i + 1$}
    \EndFor
    \\
    \State{$\texttt{search\_vec} \leftarrow bundle(\texttt{ngram}_0,
      \texttt{ngram}_{1}, ...)$}
    \State{$\texttt{idx} \leftarrow 0$}
    \State{$\texttt{min\_distance} \leftarrow \infty$}
    \State{$\texttt{class\_idx} \leftarrow 0 $}
    \For{p in prototype vectors}
    \State{$\texttt{distance} \leftarrow popcount(\texttt{search\_vec} \oplus
      p)$}
    \If{$\texttt{distance} < \texttt{min\_distance}$}
       \State{$\texttt{min\_distance} \leftarrow \texttt{distance}$}
       \State{$\texttt{class\_idx} \leftarrow \texttt{idx} $}
    \EndIf
    \EndFor
    
\end{algorithmic}
\end{listing}

Instead of recalculating the same character vectors repeatedly when
sliding over the sentence, we recursively compute the n-gram using a FIFO structure \cite{Joshi2017}.
Mapping the above algorithm to the proposed ISA with an AM size
of 16 vectors and vector fold of one results in the following code:

\begin{listing}[htbp]
\label{lst:1}
\begin{minted}[escapeinside=||, linenos, fontsize=\scriptsize, numbersep=-5pt, tabsize=4]{asm}
	start:
	  hw.loop0 nr_characters_in_sentence, end_loop
	  enc_reg |$\rightarrow$| mix |$\rightarrow$| enc_reg
	  mem[12] |$\rightarrow$| mix |$\rightarrow$| bind_with_enc_reg |$\rightarrow$| mem[11]
	  mem[13] |$\rightarrow$| mix |$\rightarrow$| mem[12]
	  mem[14] |$\rightarrow$| mix |$\rightarrow$| mem[13]
	  mem[15] |$\rightarrow$| mix |$\rightarrow$| mem[14]
	  # Generate seed for char vector
	  zero_vec |$\rightarrow$| man |$50\%$| |$\rightarrow$| enc_reg
	  #IM-Map char represented with 5-bits
	  MIX_EXT 5 #5+2 cycles
	  enc_reg |$\rightarrow$| mem[15]
	  mem[11] |$\rightarrow$| bind_with_enc_reg |$\rightarrow$| bundle
	end_loop:
	  threshold_bndl_cntrs |$\rightarrow$| mem[15]
	  am_search nr_classes #nr_classes+2 cycles
	  intr 400, 2
	  jmp start
\end{minted}
  \caption{Microcode mapping of the language classification algorithm in pseudo
code.
Arrows indicate that operations happen in a combinational pipeline in the same cycle,
multi-cycle instructions are specially indicated with comments denoting the
number of execution cycles}
\end{listing}
We omitted the initialization steps that would correspond to lines 1-3 in the pseudo-code listing for simplicity.
As can be seen in listing \ref{lst:1}, the body of the algorithm maps to the 12 instructions (lines 1-16).
The instruction on line 17 triggers an interrupt if the processed sentence belongs to the classes represented by prototype 1 or 2 with a Hamming distance of less or equal to 400 bits.
The final unconditional jump causes the algorithm to start over again, either immediately if the interrupt conditions are not met or after the host processor clears the pending interrupt.

\section{Implementation and Results}
\label{sec:impl-results}
In this section, we evaluate the proposed architecture in terms of area and power consumption.
In Section~\ref{sec:overh-analys-scm}, we present an overhead analysis of the proposed associative memory.
Finally, in subsection \ref{sec:tuning-maxim-energy} we compare the area and power consumption of the whole accelerator for two different technologies nodes and examine the influence of the vector fold parameter on the efficiency for a given target technology.
\subsection{Methodology}
\label{sec:methodology}
We followed the subsequent methodology for the area and power analysis; the purely digital design written in SystemVerilog RTL was first synthesized with Synopsys Design Compiler 2018.6 using default settings for mapping effort.
We evaluate the design's performance in two different target technologies: The first one is a 65\,\si{\nano\meter} Low-Leakage Low-K process node using a high Vth (HVT) standard cell library to minimize cell leakage at low operating frequencies required by the HDC accelerator.
If not denoted otherwise, all numbers were obtained with the typical case library characterization at 1.0\,\si{V}, 25\,\textdegree C.
The second technology we targeted is a 22nm FDSOI node using a UHVT and SLVT library.
 \todo{Verify that degree symbol is rendered correctly.}
The library characterization at 0.8\,V, 25\, °C without body biasing at the typical-typical corner was used.
Using Cadence Innovus 2018, we performed place and route with an eight-layer metal stack for the 65\,nm node targeting a core area utilization of 80\%. For the 22\,nm node, a ten-layer metal stack with a target core area utilization of 70\% was used.
Post-layout power numbers were obtained with Cadence Voltus using switching activity for all internal nodes extracted from a timing back-annotated post-layout simulation of the HDC algorithms in Mentor Graphics Questasim 2019.

\subsection{Energy and Area overhead Analysis of SCM based AMs}
\label{sec:overh-analys-scm}
Table \ref{tab:scm_am_energy_efficiency} provides an evaluation of the area overhead and energy efficiency for a fully-combinational and the row-sequential AM architecture described in Section~\ref{sec:synthesizable-cam}.
To get an accurate estimate of the delay and power consumption at sub nominal voltages, the complete standard cell library was recharacterized with spice simulations using Cadence Liberate for a $V_{DD}$ corner of 0.6\,\si{V}.
At this voltage, all standard cells within the library are still operational in spice simulation.

6T-bitcell based SRAMs that are readily available in all commercial technology nodes are no longer operational at such low voltages\cite{Meinerzhagen2011, Sinangil2008}.
Although there are specialized low-voltage SRAMs for sub-threshold operation \cite{Mohammadi2018}, they are custom-tailored for a particular technology and not readily available for all technology nodes. Furthermore, experiments by \citeauthor{Andersson2016a} indicate that customized SCMs can still have an energy advantage over sub-threshold SRAMs for small memory sizes \cite{Andersson2016a}.

At the 0.6V operating corner, we see a 4$\times$ improvement in energy efficiency for the sequential architecture and almost 5$\times$ for the fully parallel version compared to operation at nominal voltage. The full-parallel implementation is 2.6x more energy efficient than the sequential one.
However, for most HDC algorithms, the vast majority of the proposed HDC accelerator's compute time is spent on vector encoding, during which the AM lookup logic stays idle.
For this reason, we focus on the row-sequential SCM AM architecture, which has a better trade-off between energy efficiency during lookup operation and static leakage power in the subsequent analysis.

\begin{table*}
  \label{tab:scm_am_energy_efficiency}
  \centering
  \begin{tabular}{llcSScSScSS}
    \toprule
    & Area [kGE] & \phantom{abc} & \multicolumn{2}{c}{Throughput [MOPS/s]} & \phantom{abc} & \multicolumn{2}{c}{Energy Efficiency [\si{\pico\joule/lookup}]} & \phantom{abc} & \multicolumn{2}{c}{Leakage Power [uW]}\\
    \cmidrule{4-5} \cmidrule{7-8} \cmidrule{10-11}
    &     & & @1.2\,\si{V} & @0.6\,\si{V}                 & & @1.2\,\si{V} & @0.6\,\si{V} & & @1.2\,\si{V} & @0.6\,\si{V}\\
    \midrule
    SRAM + Digital AM    & 17  & & 2.56  & \multicolumn{1}{c}{---} & & 3280  & \multicolumn{1}{c}{---} & & 1.5 & \multicolumn{1}{c}{---}\\
    Sequential SCM AM    & 101 & & 1.29  & 0.23                  & & 2353  & 556 & & 7.5 & 1.7\\
    Full parallel SCM AM & 265 & & 13.80 & 1.54                  & & 921   & 188 & & 81.0 & 15.1\\
    \bottomrule\\
  \end{tabular}
  \caption{Area and Energy Efficiency comparison of SCM based 128 by 128 bit AM- and SRAM based AM-Architecture in 65nm technology using all three available VT flavors.
The most energy efficient SRAM configuration generated by the available SRAM macro generator collection for the target technology was chosen.
}
\end{table*}


\subsection{Tuning for Maximum Energy-Efficiency}
\label{sec:tuning-maxim-energy}
\begin{figure*}[h]
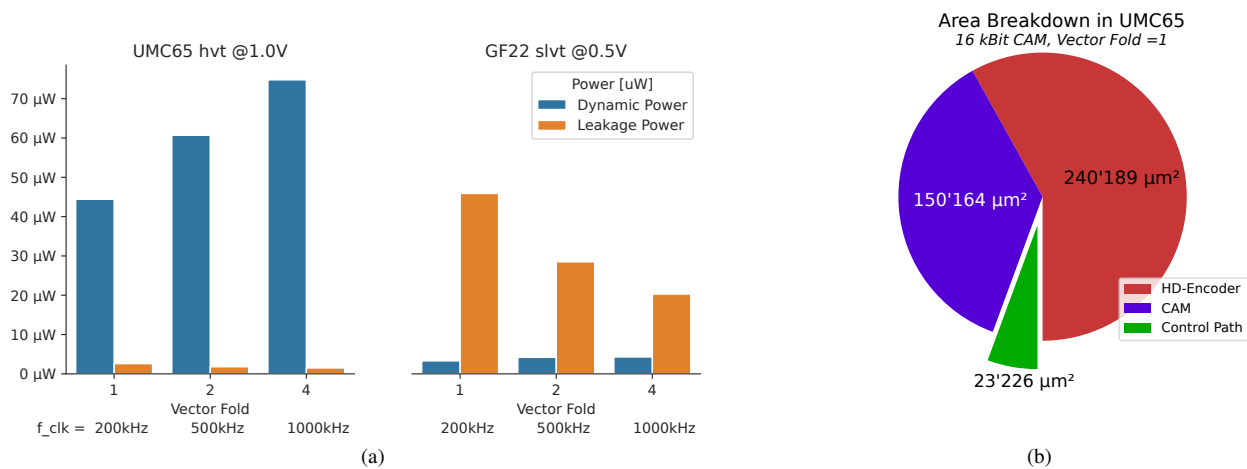

  \centering
  \subfloat[]{
    \includesvg[width=0.6\linewidth]{figs/folding_vs_power}
    \label{fig:folding_vs_power}
  }
  \subfloat[]{
    \includesvg[width=0.35\linewidth]{figs/area_breakdown_hd_accel.svg}
     \label{fig:area_breakdown_umc65}
  }
  \caption{\protect\subref{fig:folding_vs_power} Post-layout-simulated power consumption of the HDC accelerator (16 vectors à 1024 bits) when executing a realtime HDC algorithm for different vector folds in 65nm and 22nm technology.
    \protect\subref{fig:area_breakdown_umc65} Area breakdown of the
    HDC algorithm for a vector fold of 1, placed and routed in UMC 65nm.}
  \label{fig:umc65_breakdwown}

\end{figure*}

As will be further elaborated in Section~\ref{sec:appl-comp}, the high amount of parallelism in the datapath and the efficiency of the proposed ISA in executing common HD-algorithms allows the architecture to be clocked at fairly low frequencies while still achieving real-time processing capabilities for many target applications.
Figure \ref{fig:folding_vs_power} shows the power breakdown of the proposed architecture synthesized with an AM size of 16kBit (16x 1024 bits) while processing an EMG gesture recognition classification algorithm for different degrees of vector folding.
Since higher vector fold values result in less datapath parallelism, we adjusted the frequency for each different vector fold configuration to achieve identical throughput for all configurations.
In other words, although the different configurations run at different frequencies, they perform the same amount of useful work per time interval with different degrees of sequentiality.

We see entirely orthogonal tendencies for the two different technology nodes in energy efficiency versus Vector Fold.
For 65nm, the overall energy efficiency increases with lower vector folds, thus a higher degree of parallelism, while we see the opposite effect in GF22.

The reason behind this effect becomes evident when we have a closer look at the area breakdown in figure \ref{fig:area_breakdown_umc65}. For a Vector Fold value of one, almost 60\% of the accelerator area is occupied by the HD-Encoder.
In a technology node like GF22 with SLVT cells, the design is dominated by leakage power.
Increasing the vector fold that directly affects the encoder's datapath width has a large effect on the overall area and thus static current draw of the accelerator.

Although the fully synthesizable architecture's technology independence would make it easy to switch to a different technology node with lower leakage, this is not always a possibility, especially when the device is integrated into a larger system.
For these situations, the vector fold feature, in addition to its function as a control knob to trade-off area for maximum throughput, provides the means to tune the design for maximum energy efficiency depending on the target technologies' leakage behavior.





\section{Applications and Use Cases}
\label{sec:appl-comp}
As thoroughly discussed in Section~\ref{sec:progr-hdc-accel}, the proposed HDC accelerator uses hardware-friendly embodiments of commonly used HDC primitives and combines them with a programmable control path.
In this section, we take a closer look at the achieved accuracy of the proposed architecture when configured to execute different classification problems using state-of-the-art HDC algorithms.
Both, to validate the soundness of the algorithmic transformations and to compare the energy efficiency with other fully digital HDC accelerators.


\subsection{Accuracy Analysis on Text classification and EMG Gesture Regocnition}
\label{sec:accur-analys-text}
\begin{figure}[t]
  \centering
  \includesvg[width=0.9\linewidth]{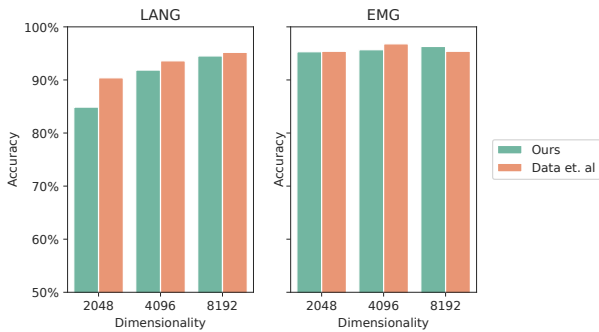}
  \caption{Achieved accuracies for the target applications using different HD-Vector sizes.}
  \label{fig:accuracy_vs_dim_plot}
\end{figure}

As mentioned earlier, the language classification of textual data is a prime example for classification with HDC.
While this application does not fit the context of always-on smart sensing, it serves the purpose of validating the accuracy implications of the permutation-based item memory materialization described in Section~\ref{sec:mixing-stage}.
We tackle the same classification task to classify the text samples into 21 Indo-European languages~\cite{Rahimi2016a}.
We use the same HDC algorithm described in Section~\ref{sec:an-isa-hd} with an n-gram size of five, which is identical to the algorithm used by \citeauthor{Rahimi2016a}.
Figure \ref{fig:accuracy_vs_dim_plot} indicates the achieved accuracy using a vector fold factor of 1 for different dimensionalities; For 8192 bit HD vectors, the modified HDC operators achieve an accuracy of 94.52\%.
This accuracy is almost identical to the results reported by \citeauthor{Datta2019} on their accelerator (95.2\%) \cite{Datta2019}.
The algorithm maps to only 14 HDC ISA instructions and has a memory requirement of five vector slots in the AM, in addition to the 21 language prototype vectors, for intermediate results during the encoding process.
For a vector fold of 1, the algorithm executes at 14 cycles per processed input character, which results in ~1400 cycles to classify a single sentence.


The second application we evaluate is hand gesture recognition on electromyography (EMG) data recorded on the subject's forearm.
We used the dataset and preprocessing pipeline from \cite{Moin2018}; The data consists of recordings from the subject performing five different hand gestures captured by a 64-channel EMG sensor array with a sampling rate of 1kSPS.
The actual HDC classification algorithm works as follows; For each time sample, the 64 channel values are continuously mapped to HD-vectors using the similarity manipulator module described in Section~\ref{sec:simil-manip} and bound to a per-channel label vector, generated in the mixer stage.
Bundling the resulting 64 channel vectors together yields a single HD-vector that represents the state of all channels for a given instance in time.
Five of these vectors are combined to a 5-gram analog to the language classification algorithm to form the search vector for associative lookup against the prototype vector.
Training of the prototype vectors works like classification, but many search vectors corresponding to the same gesture are bundled together to form the prototype vector.

The whole algorithm maps very well to HDC ISA, requiring only 12 instructions and two memory slots for intermediate results.
The inner loop over the 64 channels in the algorithm is executed in only two cycles for a folding factor of 1, which results in a total of \textbf{678 cycles} to classify a single 500ms window of data. Consequently, realtime classification of 64 EMG channels implies an accelerator clock frequency of only \textbf{1'356 Hz}.

While the data preprocessing flow we used in our experiments was identical to \cite{Moin2018}, the HDC algorithm, although identical in general structure, differs in a few crucial aspects from the baseline implementation.
 \citeauthor{Moin2018} perform CIM mapping of the individual samples to HDC vectors using scalar multiplication of the sample value with a per-channel bipolar label vector, effectively leaving the binary domain \cite{Moin2018}.
Moreover, the bundling operation to form a time sample vector is implemented as a scalar addition of the integer-valued vectors before thresholding the result back to a bipolar representation with positive values mapped to +1 and negative values to -1.
Even though the proposed algorithm modification stays strictly in the binary domain, there is only a small drop in accuracy; With 8192 dimensions, the proposed architecture achieves 96.31\% accuracy while \citeauthor{Moin2018} report an accuracy of 99.44\% accuracy using 10'000 bit-vectors and arbitrary precision bundling \cite{Moin2018}.


\subsection{Ball Bearing anomaly detection}
\label{sec:ball-bearing-anomaly}
Predictive maintenance, also known as condition-based maintenance, is a term for the process of estimating the current condition of in-service equipment to anticipate component failure.
The goal is to switch to a maintenance scheme were components are replaced once they approach their end-of-life instead of fixed maintenance intervals based on preventive replacement according to the statistically expected lifetime \cite{Selcuk2017}.
As part of our algorithmic investigations, we investigate the feasibility of HDC for the task of ball bearing fault prediction using vibration data from low power accelerometer sensors.



For our analysis, we use the IMS Bearing Dataset provided by the University of Cincinnati \cite{Qiu2006}.
They recorded vibration data at a sampling rate of 20kHz from 4 different ball bearings on a loaded shaft rotating at a constant 2000rpm.
We concentrated on the first of the three recording sets, which contains 1 second data records obtained with an interval of 10 minutes in a run-to-failure experiment that lasted 35 days with an accumulated operating time of about 15 days.

\begin{figure}[ht]
  \centering
  \includegraphics[width=\linewidth]{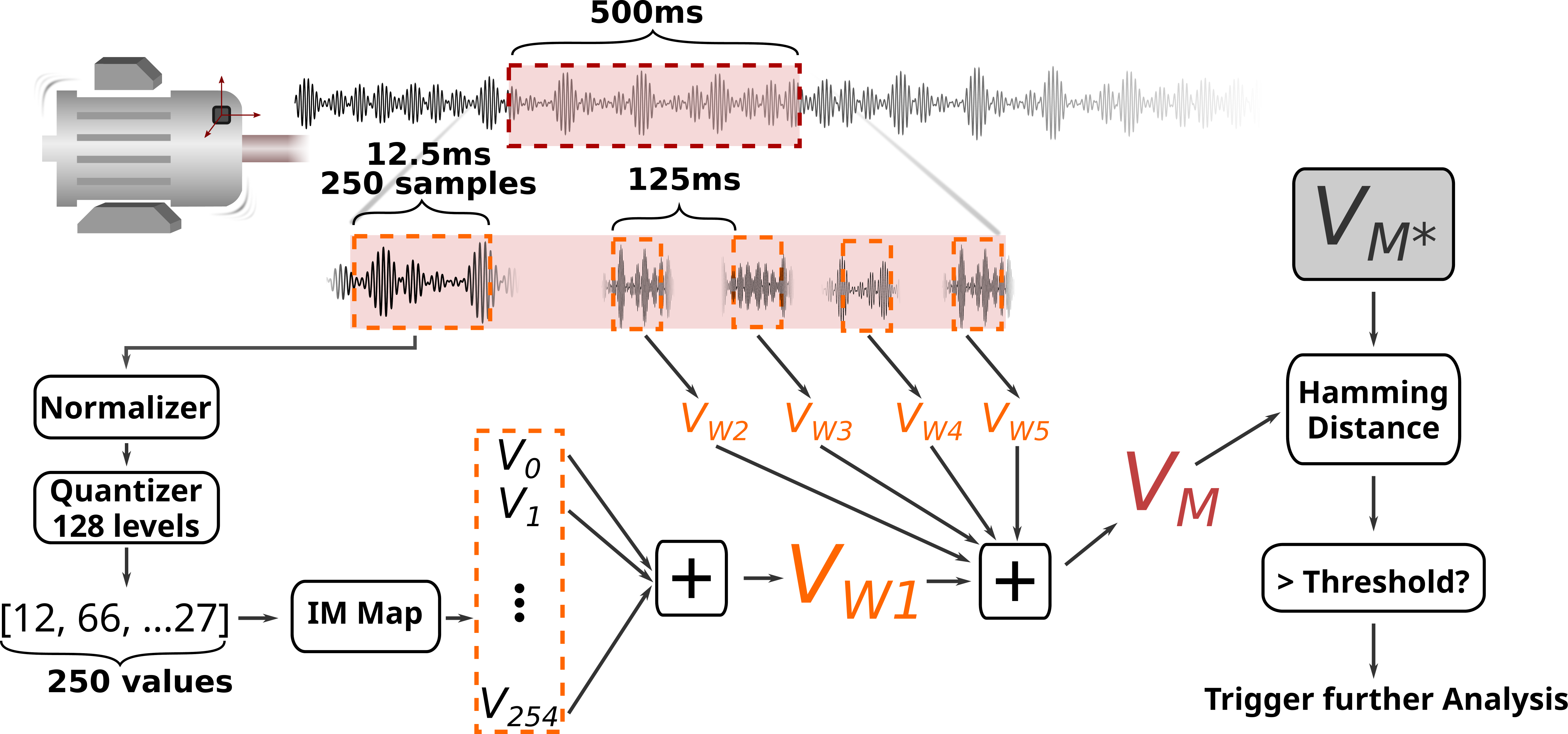}
  \caption{\label{fig:illustration_ball_bearing_algo} Illustration of the proposed HDC based ball bearing anomaly detection algorithm. $V_{M*}$ denotes the on-line trained calibration vector from the first 24 operating hours of the ball bearing.}
\end{figure}

Figure \ref{fig:illustration_ball_bearing_algo} illustrates the basic classification procedure.
The algorithm requires an initial calibration phase where a prototype vector representing the ball bearing's normal operating condition is generated.
With the inherent feature of HDC that classification and training are of almost equivalent computational complexity, online-training with HDC imposes negligible additional energy costs.
The current control path of the proposed HDC accelerator allows for online training algorithms to be encoded in the algorithm storage but requires an external control entity, e.g., a general-purpose core that provides the labels during algorithm execution.

The algorithm's basis is the encoding of small time windows from the raw vibration data to \emph{measurement vectors} $V_M$.
Each time window consists of 250 samples (12.5ms).
The sample values are first normalized using a pre-trained normalization factor and quantized to 7 bits.
Each sample value is then mapped to an HD-vector using IM mapping, and the whole window of 250 samples is bundled together to a \emph{window vector} $V_W$.
Five of these window vectors with an interval of 125ms are again bundled together to form a single \emph{measurement vector} $V_M$.
The resulting vector thus approximates the amplitude distribution over a 0.5-second time frame.

The general idea behind the proposed analysis scheme is to generate a prototype vector $V_{M*}$ using the first couple of \emph{measurement vectors} after commissioning.  We then track the evolution of Hamming distance over time for subsequent measurement vectors.
We calibrated the prototype vector using 100 random measurement vectors from the first 24 operating hours of the respective ball bearing in our experiments.
Similarly, the normalization factor is generated using the 99\% quantile of the amplitude within the same 24 hours after commissioning.
The proposed algorithm can be mapped to \textbf{9 HDC ISA instructions} and requires two vector slots, one for the calibration vector and one for intermediate results.

\begin{figure}
  \centering
  \includesvg[width=0.9\linewidth]{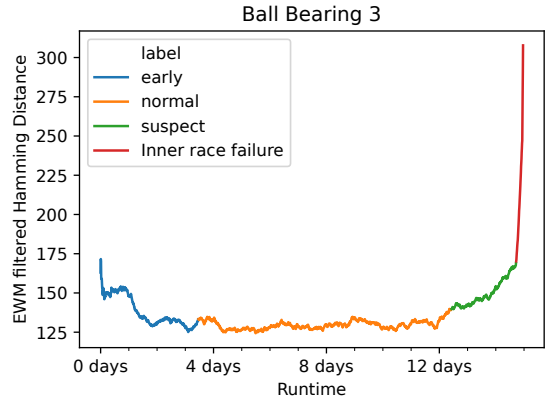}
    \caption{Hamming distance evolution over time for ball bearings 3 in the IMS dataset. The Hamming distance was post-processed with an exponential moving average filter with a halflife of 5 hours. The other ball bearings in the dataset show a similar behavior}
    \label{fig:ball_bearing_hamming_dist_plot}
\end{figure}

Figure \ref{fig:ball_bearing_hamming_dist_plot} shows the evolution of Hamming distance over time with an exponential moving average filter with a half-life of five hours. This feature can be computed very efficiently without the need for a large ring buffer.
The line color indicates the labels proposed by experts on manual analysis of the dataset \cite{BenAli2015}.

By the end of the IMS ball bearing experiment, bearings 3 and 4 failed, while bearings 1 and 2 were severely worn down but did not fail yet.
We see a sharp increase in Hamming distance for all four ball bearings several hours before the actual failure, in the case of ball bearing 3, even several days before the actual inner race failure.


While the proposed algorithm certainly does not replace more involved analysis on time and frequency domain features, the results suggest that it can act as a first filtering stage for aggressive duty cycling of more power-intensive analysis schemes when combined with simple thresholding.
However, more experiments on larger datasets and possibly with more complex HDC encoding schemes will be required to quantify the benefits of an HDC based ball bearing fault predictor.


\subsection{Energy Efficiency Analysis and Comparison}
\label{sec:accur-energy-effic}

\begin{table*}
  \centering
  \resizebox*{\linewidth}{!}{
    \begin{tabular}{lcrcrrcrrcrr}
      \toprule
      Algorithm & \phantom{a} & \# of HDC instr. & Vector Memory     & Cycles/classif. & Realtime Freq. [\si{\kilo\hertz}] &\phantom{a} & \multicolumn{2}{c}{Power [\si{\micro\watt}]} & \phantom{a} & \multicolumn{2}{c}{Min. Energy/classif. [\si{\nano\joule}] @\SI{100}{\kilo\hertz}} \\
      \cmidrule{8-9} \cmidrule{11-12}
                &             &                  &                   &                 &                                   &            & 65nm        & 22nm                            &            & 65nm        & 22nm\\
      \midrule                                                        
      LANG      &             & 14               & 5 + 21            & 1400            & 100                               &            &  86.5       &  23.7                           &            &      1205   & \textbf{332}  \\
      EMG       &             & 12               & 2 +\space\space 5 & 678             & 1.4                               &            &  10.7       &  2.9                            &            &       703   & \textbf{191} \\
      BEARING   &             & 9                & 1 +\space\space 1 & 12513           & 25                                &            &  29.1       &  7.9                            &            &      10913  & \textbf{2966} \\
    \end{tabular}
  }
  \caption{Memory requirements and post-layout energy numbers of selected HDC
    algorithm on the proposed architecture with an AM size of 32 x 2048 bit,
    vector fold 1}
  \label{tab:algorithm_energy}
\end{table*}


Table \ref{tab:algorithm_energy} summarizes the performance of the three introduced HDC algorithms, language classification (LANG), EMG gesture recognition (EMG), and ball bearing anomaly detection (BEARING).
Columns 2 \& 3 report the number of HDC instructions and the total number of required HD vector memory to map the algorithm to the architecture.
Column 4 shows the required minimum frequency for real-time execution of the algorithm (not applicable for LANG since there is no real-time constraint for this application).
The last two columns indicate the power when operating at the aforementioned minimum frequency and the corresponding energy efficiency per classification.
For LANG, we consider a single classification to be the processing of a 100-character string, the average sentence length in the Wortschatz corpora.
For EMG and BEARING, a single classification is defined as the analysis of a 500ms window as described in the algorithm sections \ref{sec:accur-analys-text} and \ref{sec:ball-bearing-anomaly}.

\begin{table*}
  \centering
  \begin{tabular}{lrrrrcrr}
    \toprule
                           & Technology & Area [kGE] & Architecture Type & IM / CIM resolution [bit] & \phantom{a} & \multicolumn{2}{c}{Energy eff.[\si{\nano\joule/inference}]}\\
    \cmidrule{7-8}
                           &            &            &                   &                           &             & LANG          & EMG \\
    \midrule                                                                                        
    \citeauthor{Datta2019} & TSMC28     &  3618      & generic           & 10 or 10                  &             & 250           & 610\\
    \textbf{Our Work}      & \textbf{GF22}       &  \textbf{1094}      & \textbf{general-purpose}   & \textbf{arbitrary and 7}           &             & \textbf{332}           & \textbf{191}\\
  \end{tabular}
  \caption{Area and Energy efficiency comparison with the current state of the art HDC accelerator architecture. The terms \emph{generic} and \emph{general purpose} were introduced by \citeauthor{Datta2019} in \cite{Datta2019}.}
  \label{tab:soa-comparison}

\end{table*}

In table \ref{tab:soa-comparison} we compare the energy efficiency of our solution to the current SoA HDC accelerator architecture from \citeauthor{Datta2019} \cite{Datta2019}.
Among other algorithms, they report the energy numbers for EMG and LANG executed on a 32 by 2048-bit accelerator in TSMC28.
We achieve a technology scaled area reduction by \textbf{3.3$\times$}.
This can be explained by massive area reductions in all major components of the accelerator. 
The most considerable effect has the on-the-fly pseudo-random materialization of the item vectors used in our design, which removes the necessity to incorporate a large ROM to store all possible item vectors. 
In fact, 62\% of the overall area in \citeauthor{Datta2019} is occupied by a large 1024 by 2048 bit ROM.
Besides the area and energy implications, the ROM based solution has the added drawback of having a hardwired partitioning of the memory; One for the item memory, containing quasi-orthogonal vectors, and one for continuous item memory vectors, where the pair-wise Hamming distance between the vectors correlates to the difference of the corresponding input values.
Another large reduction in area is achieved in the AM, where our solution uses latch cells and sequentially calculates the Hamming distance in contrast to the baseline, which uses a flip-flop based fully parallel implementation.

In fairness, one has to notice that \cite{Datta2019}, with a maximum clock frequency of 434MHz, unarguably has a much higher peak throughput than our solution due to its parallel and heavily pipelined architecture.
However, the results in table \ref{tab:algorithm_energy} suggest that algorithms used for always-on sensing do not benefit from such a high throughput, and energy efficiency is the key metric by which we should judge the performance of the different approaches.

As we can see in table \ref{tab:soa-comparison}, the energy efficiency differences between the two architectures depend a lot on the algorithm at hand.
For LANG, the achieved energy efficiency is slightly worse (+31\%) than the baseline, which is still impressive considering the 3.3$\times$ reduction in area.

For EMG, on the other hand, we achieve a 3.1$\times$ improvement in energy efficiency.
This can be explained by the difference in the computational complexity of orthogonal and continuous item mapping in our architecture.
In LANG, input values are mapped to quasi-orthogonal vectors using the mixing stage (\ref{sec:mixing-stage}), which requires $log_2(N)$ cycles, where $N$ denotes the cardinality of the input set.
The overhead of this iterative approach considerably lowers the energy advantage of not using a large ROM for item memory generation.
For EMG, on the other hand, the input values are mapped continuously using the similarity manipulator, which can be performed in a single cycle and can even be combined with a bundling or bin operation in the subsequent encoder units.
Hence, for this algorithm, the effect of not requiring a ROM comes to display.
In general, we can say that for very high input value resolutions, the overhead of iterative item vector generation starts to dominate the overall energy consumption of our architecture.
Still, the fact that the computational complexity of the rematerialization approach grows with the logarithm of the input space instead of linear ROM area scaling suggests an advantage of our architecture for larger input space cardinality.
In any case, the proposed architecture excels in its energy proportionality to the desired HDC algorithm.
The ROM based approach in \cite{Datta2019} has an almost fixed cost for item memory mapping with an upper limit on the supported resolution.
For example, in LANG, only 13\% (27 out of 1024 item vectors) of all ROM entries are required to map the input values.
The architecture proposed by \citeauthor{Datta2019} is only \emph{generic} according to their taxonomy on HDC algorithm classes \cite{Datta2019}. 
In contrast, the microcode based approach that our architecture follows allows for arbitrary HDC algorithm computation within the limits of the available AM and instruction memory resources.
Finally, our proposed architecture is energy- and area-flexible and can be finely parametrized to fit the area, throughput, and energy efficiency constraints of a particular target technology.



\section{Conclusion}
\label{sec:conclusion}
In this work, we presented a novel all-digital cross-technology mappable HDC accelerator architecture with a highly configurable datapath using a newly proposed microcode ISA optimized for HDC.
Place and routed in GF 22nm technology, the architecture improves on the current state of the art both in area and energy efficiency by a factor of up to 3.1$\times$ and 3.3$\times$ respectively.
The architecture achieves an energy efficiency of 192 \si{\nano\joule/inference} for the task of EMG gesture classification with an always-on compatible typical power consumption of ~5\,\si{\micro\watt}.
Our post-layout simulation experiments on different digital associative memory architectures in Section~\ref{sec:overh-analys-scm} indicate a significant potential for latch based associative memories to push the limits of energy efficiency when operating at sub-nominal voltage and can already outperform the energy efficiency of commercial-off-the-shelf SRAM macros at nominal voltage.
In Section~\ref{sec:appl-comp} we demonstrated that our newly introduced rematerialization scheme for IM and CIM mapping have a negligible impact on classification accuracy with a drop of less than 0.5\% compared to a ROM based approach used by the current SoA HDC accelerator.
As part of the analysis, we proposed a novel HDC based end-to-end classification algorithm for ball bearing anomaly detection that maps to only 9 HDC microcode instructions.
While our experiments in Section~\ref{sec:accur-energy-effic} indicated that the energy efficiency of a rematerializing IM is inferior to a ROM based solution for low input resolutions, the proposed CIM mapping scheme outperforms the current SoA in energy efficiency, area usage, and flexibility.
Finally, we provided the first open-source release of a complete HDC Accelerator platform which is possible due to the all-digital nature of the proposed architecture.

\bibliography{references_manuel,bst_control} \bibliographystyle{IEEEtranN}

\end{document}